\documentstyle[preprint,aps]{revtex}
 
\def\be{\begin{equation}}
\def\ee{\end{equation}}
\newcommand{\bea}{\begin{eqnarray}}
\newcommand{\eea}{\end{eqnarray}}
\def\we{\approx}
\def\fr{\frac}
\def\a{\alpha}
\def\b{\beta}

\def\e{\epsilon}

\def\l{\lambda}
\def\m{\mu}
\def\n{\nu}
\def\r{\rho}
\def\s{\sigma}

\def\th{\theta}
\def\w{\omega}
\def\W{\Omega}

\def\L{\Lambda}
\def\CL{\cal L}
\def\p{\partial}
\begin{document}
\draft
\title{{Topologically Massive Abelian Gauge Theory From BFT Hamiltonian 
Embedding of A First-order Theory}}
\author{E. Harikumar  and M. Sivakumar$^{**}$}
\address{ School of Physics \\
 University of Hyderabad\\
 Hyderabad, Andhra Pradesh\\
 500 046  INDIA.}
\footnotetext {$^{**}$ Electronic address: mssp@uohyd.ernet.in }  
\maketitle

\begin{abstract} 
We start with a new first order gauge non-invariant formulation of massive 
spin-one theory and map it to a reducible gauge theory viz; abelian 
$B{\wedge}F$ theory by the Hamiltonian embedding procedure of Batalin, 
Fradkin and Tyutin(BFT). This equivalence is shown from the equations of 
motion of the  embedded Hamiltonian. We also demonstrate that the original 
gauge non-invariant model and the topologically massive gauge theory can both 
be obtained by suitable choice of gauges, from the phase space partition 
function of the emebedded Hamiltonian, proving their equivalence. Comparison 
of the first order formulation with the other known massive spin-one theories 
is also discussed.

\end{abstract}

\draft
\newpage
{\large I. Introduction}

The construction and the study of massive spin-1 theories which are also gauge
invariant has a long history \cite{js}, the well-known example being models
with Higgs mechanism. Since the existence of Higgs particle has not yet been
experimentally verified, it prompts a closer look at other models wherein
mass and gauge invariance coexist. One such model which is currently being
studied is the topological mechanism for gauge invariant mass for spin-one
particle without a residual scalar field, wherein vector and tensor (2-form) fields are coupled in a gauge 
invariant way by a term known as $B{\wedge}F$ term \cite{bf}. The Lagrangian for 
this model (for Abelian case) is given by
\be
{\CL} = -\fr{1}{4} F_{\m\n}F^{\m\n} + \fr{1}{2\cdot 3!} H_{\m\n\l}H^{\m\n\l}
+\fr{1}{4}\e_{\m\n\l\s} B^{\m\n}F^{\l\s} ,
\ee
\noindent where $H_{\m\n\l}= \p_\m B_{\n\l}~+~{\rm cyclic~ terms}.$
This Lagrangian is invariant under
\bea
 A_\m \rightarrow A_\m + \p_\m \l,\\
 B_{\m\n}\rightarrow B_{\m\n} + (\p_\m \l_\n - \p_\n \l_\m).
\eea
\noindent A similar construction has also been made for non-Abelian
theory \cite{tma}.

The invariance of $B_{\m\n}$ when $\l_\m =~\p_\m \w,$ in (3) necessitates 
the introduction of ghost for ghost terms in BRST quantization of the 
$B{\wedge}F$ theory \cite{al}. It can also be seen in the constraint 
quantization where this invariance makes the generators of gauge 
transformation linearly dependent \cite{neto}. This theory shows a similar constraint 
structure as that of massless Kalb-Ramond theory in the existence of first 
stage reducible constraint \cite{rkrb}.

The purpose of this paper is to construct a new  first-order formulation 
of massive spin-one theory
involving vector and 2-form fields (see eqn (8) below), which is {\it gauge 
non-invariant} and by following the idea of Hamiltonian embedding 
due to Batalin, Fradkin and Tyutin (BFT) \cite{bft}, 
we show that the
resulting theory is equivalent to $B{\wedge}F$ theory (1).
Thus we have a novel representation of the abelian $B{\wedge}F$ theory in
terms of a gauge non-invariant first order formulation.

The motivation for this study is two fold. One is that the $B{\wedge}F$ 
theory, in addition to being a candidate as an alternate for  
Higgs mechanism also appears in diverse areas like condensed matter
physics \cite{ps} and black-holes \cite{fw}. Hence a 
model which is an
equivalent realization of it has potential applications. The other is
that the Hamiltionian emebedding procedure, employed here, by itself
is of current interest. Several models like abelian and non-abelian 
self-dual model in $2+1$
dimensions, abelian and non-abelian Proca theories have been studied 
in detail in recent times applying the BFT procedure \cite{rb}. Here
we apply it, along the lines of
\cite{rb1} to demonstrate the equivalence between self-dual model 
and Maxwell-Chern-Simons theory, to establish the equivalence of this 
gauge non-invariant theory (8)to the {\it reducible } gauge theory (1).

\noindent We work with $g_{\m\n}= \rm{diag}~(1 -1 -1 -1)$ and $\e_{0 1 2 3}=1$

	In order to compare our formulation with the other known formulations 
of massive spin-1 theories, we first recollect them. The earliest
formulation is in the form of a first order relativistic wave equation due to
Duffin-Kemmer -Petiau(DKP) \cite{dkp}, which is given by
\be
		(i\b^{\m}\p_{\m} + m )\psi =0~,
\ee

\noindent where $\psi$ is a $10$ dimensional column vector and ${\b_\m}$ 
are $10{\times}10$ hermitian matrices obeying,
\be
{\b_\m}{\b_\n}{\b_\l} + {\b_\l}{\b_\n}{\b_\m} = g_{\m\n}{\b_\l} + g_{\l\n}
{\b_\m}.	
\ee
The other well-known theory is that of Proca Lagrangian,
\be
{\CL} = - \fr{1}{4} F_{\m\n} F^{\m\n} +\fr{m^2}{2}A_{\m}A^{\m}
\ee
There is, another formulation involving 2-form field 
 \cite{tp} given by
\be
{\CL} = \fr{1}{12}H_{\m\n\r}H^{\m\n\r} + \fr{m ^2}{4}B_{\m\n}B^{\m\n},
\ee
where $H_{\m\n\r}= \p_{\m}B_{\n\r} + {\rm cyclic~ terms}.$

Both these formulations (6, 7) can be made gauge invariant by adding 
suitable St\"uckelberg fields to compensate for the gauge variations of the mass
terms, the compensating fields being scalar and vector fields for the 
Lagrangians
(6 and 7) respectively \cite{sb}. It should be pointed out that the 
St\"uckelberg
formulations of both these theories are  equivalent by
duality transformation to $B{\wedge}F$ theory \cite{we}. 

The new first order Lagrangian describing massive spin-1 theory involving a
one form and a two form fields, which is proposed and studied in this paper, 
is given by
\be
{\CL} =- \fr{1}{4} H_{\m\n} H^{\m\n} + \fr{1}{2} G_{\m}G^{\m} + \fr{1}{2m}
\e_{\m\n\l\s}H^{\m\n}{\p^\l}{G^\s}.
\ee

This Lagrangian obviously has no gauge-invariance.
The field equations following from this Lagrangian are
\bea
-H_{\m\n} + \fr{1}{m}\e_{\m\n\l\s}{\p^\l}{G^\s}=0,\\
{\rm and}~~~~G_\m + {\fr{1}{2m}}\e_{\m\n\l\s}{\p^\n}H^{\l\s}=0.
\eea
From these equations, it follows that  
\bea
\p_\m H^{\m\n} = 0,\\
{\rm and}~~~~\p_\m G^\m= 0.
\eea
The fact that the above Lagrangian describes a massive spin$-1$ theory 
can be
easly seen by rewriting the coupled equations of motion (using the 
conditions (11) and (12)) as
\be
(\Box + m^2)H_{\m\n}=0,
\ee
\noindent or alternatively
\be
(\Box + m^2)G_\m=0.
\ee
\noindent
 Since the equation of motion (14) along with the constraint (12) follows
from Proca Lagrangian, we should expect the latter to emerge from the
above Lagrangian (8).
Indeed by integrating out $H_{\m\n}$  from the
Lagrangian (8), Proca Lagrangian (6) is obtained. Similarly, by 
eliminating $G_\m$ from the Lagrangian (8)
we arrive at the 
 Lagrangian (7). It is natural to ask how this 
Lagrangain (8) is different	
from the first-order formulation of Lagrangians (6, 7).

 The standard first-order form of Proca Lagrangian is given by
\be
{\CL}= \fr{1}{4}B_{\m\n}B^{\m\n}+\fr{1}{2}B_{\m\n}F^{\m\n}+m^2 A_\m A^\m.
\ee
Here, by eliminating the linearzing field  $B_{\m\n}$, we get back to 
the Proca Lagrangian (6). But eliminating $A_\m$ will not lead to the 
Lagrangian (7) for 2-form fields.
Similarly, the standard first order form corresponding to the 
Lagrangian (7) is
\be
{\CL}= -\fr{1}{2\cdot 3!}C_{\m\n\l}C^{\m\n\l} -\fr{1}{3!}C_{\m\n\l}H^{\m\n\l}
+ m^2 B_{\m\n}B^{\m\n}.
\ee
\noindent 
Here, too, eliminating $B_{\m\n}$ from the above Lagrangian 
will not lead to the Proca Lagrangian involving 
1-form (6).

It should be stressed that (8) is different from the standard first-order
formulations (15, 16), by being the first-order formulation for both (6)
and (7). 
 
 The first order field equations (9) and (10) can be rewritten as
\be
\left({\b_\m}{\p^\m} + m\right)\psi =0~,
\ee
\noindent where $\psi$ is a $10$ dimensional column vector whose 
elements are the independent components of $G_\m$ and $H_{\m\n}$. 
But here the ${\b_\m}$ matrices are {\it not} found to obey the 
DKP algebra (5). 

This paper is organised as follows: In section II, Hamiltonian
embedding of the Lagrangian (8) is constructed along the lines of BFT and
the embedded Hamiltonian is shown to be equivalent to that of $B{\wedge}F$
theory. Section III shows the equivalence in phase space path integral 
approach using the embedded Hamiltonian. Finally we end up with conclusion.
\vskip0.50cm
{\large II. Hamiltonian Embedding }
\vskip0.50cm
We start with the Lagrangian (8), with the last term expresed in a 
symmetric form as
\be
{\CL} =- \fr{1}{4} H_{\m\n} H^{\m\n} + \fr{1}{2} G_{\m}G^{\m} + \fr{1}{4m}
\e_{\m\n\l\s}H^{\m\n}{\p^\l}{G^\s} - \fr{1}{4m}\e_{\m\n\l\s}\p^\m H^{\n\l} G^\s.
\ee
\noindent This Lagrangian can be re written as
\be
{\CL}= \fr{1}{4m}\e_{0ijk}H^{ij}\p^0 G^k - 
\fr{1}{4m}\e_{oijk}\p^0 H^{ij}G^k~-~{\cal H}_c
\ee

\noindent where ${\cal H}_c$, the Hamiltonian density following from the 
above Lagrangian (19) is
\be
{\cal H}_c =\fr{1}{4}H_{ij}H^{ij} -\fr{1}{2}G_i G^i + H_{0i}
\left(  \fr{1}{2}H^{0i}-\fr{1}{m} \e^{0ijk}{\p_j} G_k\right)
-\fr{1}{2} G_0 \left(G^0 +\fr{1}{m} \e^{0ijk}{\p_i}H_{jk}\right),
\ee
The primary constraints are 
\bea
\Pi_0 \we 0,\\
\Pi_{0i}\we 0,\\
\W_i \equiv \left( \Pi_i -\fr{1}{4m}\e_{0ijk}H^{jk} \right)\we 0,\\  
\L_{ij} \equiv \left ( \Pi_{ij}+\fr{1}{2m} \e_{0ijk} G^k \right) \we 0,
\eea
\noindent The persistence of the primary constraints leads to secondary 
constraints,
\bea
\L \equiv \left( G_0 + \fr{1}{2m}\e_{0ijk}{\p^i}H^{jk} \right) \we 0,\\
\L_i \equiv \left(-H_{oi} + \fr{1}{m}\e_{0ijk}{\p^j}G^k \right) \we 0.
\eea
The non-vanishing poisson brackets between these linearly independent 
constraints are
\bea
\left\{ \Pi_{0}({\vec x}) , \L({\vec y})\right\} = -\delta({\vec x -\vec y}),\\
\left\{ \Pi_{0i}({\vec x}) , \L^{j}({\vec y})\right\}= \delta^{j}_{i} \delta({\vec x -\vec y}),\\
\left\{ \W_{i}({\vec x}) , \L_{j}({\vec y})\right\} = \fr{1}{m}\e_{0ijk}\p^k\delta({\vec x -\vec y}),\\
\left\{ \W_{i}({\vec x}) ,\L_{jk}({\vec y})\right\} = -\fr{1}{m}\e_{oijk}\delta({\vec x -\vec y}),\\
\left\{ \L_{ij}({\vec x}), \L({\vec y})\right\} = \fr{1}{m}\e_{oijk}\p^k\delta({\vec x -\vec y}).
\eea
Thus all the constraints are second class as expected of a theory without
any gauge invariance. Note that the constraints $\W_i~~{\rm and}~~ \L_{ij}$
are due to the symplectic structure of the Lagrangian (8). Following Fadeev
and Jackiw \cite{fj}, the symplectic conditions, which are not true 
constraints, are implemented strongly leading to the
the modified bracket,
\be
\left\{ G_{i}({\vec x}),H_{jk}({\vec y})\right\}=-m\e_{0ijk}
\delta({\vec x -\vec y}).
\ee
\noindent Consequently $\W_i$ and $\L_{ij}$ are implemented strongly.

Now we enlarge the phase space by introducing canonically conjugate
auxiliary pairs $(\a, \Pi_\a,p_i,~{\rm and}~q_i)$ and modify
the remaining second class constraints such that they are in
strong involution, i.e., have vanishing Poisson brackets. To this end, 
we define the non-vanishing Poisson
brackets among the new phase space variables to be
\bea
\left\{\a({\vec x}), \Pi_{\a}({\vec y})\right\}=\delta({\vec x -\vec y}),\\
\left\{q^{j}({\vec x}), p_{i}({\vec y})\right\}=\delta^j_i\delta({\vec x -\vec y}).
\eea
The modified constraints which are in strong involution read
\bea
\w = \Pi_0 +{\a} \,\,,\\
\L^{\prime}={\L}+ \pi_{\a} ,\\
\th_i = \Pi_{0i}+q_i\,\\
\L_{i}^{\prime} = \L_i - p_i.
\eea
Following the general BFT procedure we construct the Hamiltonion which is
weakly gauge invariant and is given by
\be
H_{GI}=\int d^3x \left[ {\cal H}_{c} + \fr{1}{2}{\Pi_{a}}^2 + 
{\a}{\p^i}G_i - \fr{1}{2}({\p^i}{\a})(\p_i{\a}) -\fr{1}{2} {p_i}{p^i} - 
\fr{1}{2}q_{ij}H^{ij} +\fr{1}{4} q_{ij}q^{ij}\right],
\ee
\noindent where $q_{ij} = (\p_{i} q_{j} - \p_{j}q_{i}).$
The  Poisson brackets of modified constraints with $H_{GI}$ are
\bea
\left\{ \w ,H_{GI}\right\} = \L^{\prime},\\
\left\{ \L^{\prime} , H_{GI}\right\} = 0,\\
\left\{ \th_i ,H_{GI}\right\} = \L_i^{\prime},\\
\left\{ \L_i^{\prime}, H_{GI}\right\} = 0.
\eea
Thus all the modified constraints are in involution with the $H_{GI}$
as one can easily see from their Poisson brackets. The gauge
transformations generated by these first class constraints (35 to 38) are 
\bea
\left \{ G^0 , \int {d^3x}\w \tilde{\th} \right\} = \tilde{\th},\nonumber\\
\left \{ \Pi_{a} , \int {d^3x}\w \tilde{\th} \right\} = -\tilde{\th},\nonumber\\
\left \{ H_{0i} , \int {d^3x}{\th_i}\psi^i \right\} = \psi_i,\nonumber\\
\left \{ p_i ,\int {d^3x}{\th_i}\psi^i \right\} = -\psi_i,\nonumber\\
\left \{ G_i ,\int {d^3x} \L^{\prime} \tilde{\th} \right\} = -\p_i\tilde{\th},\nonumber\\
\left \{{\a}, \int {d^3x}\L^{\prime} \tilde{\th} \right\} = \tilde{\th},\nonumber\\
\left \{ H_{ij}, \int{d^3x}\L^{\prime}_i\psi^i \right\} = -(\p_i\psi_j - \p_j\psi_i),\nonumber\\
\left \{ q_i, \int{d^3x}\L^{\prime}_j\psi^j \right\}=-\psi_i.
\eea
Thus the combinations
\bea
\bar{G}_0 = G_0 + \Pi_{a},\nonumber\\
\bar{G}_i= G_i+\p_i\a,\nonumber\\
\bar{H}_{oi} = H_{oi} + p_i,\nonumber\\
\bar{H}_{ij}= H_{ij} -~q_{ij}.
\eea
\noindent are gauge invariant under the transformation generated
by the first class constraints (35 to 38). 
Next we re-express gauge invariant Hamiltonian density ${\cal H_{GI}}$ in
terms of these gauge invariant combinations,
\be
{\cal H_{GI}} = \fr{1}{4} \bar{H}_{ij}\bar{H}^{ij} 
-\fr{1}{2}\bar{H}_{oi}\bar{H}^{oi}-\fr{1}{2}\bar{G}_i\bar{G}^i
+\fr{1}{2}\bar{G}_0\bar{G}^0
 -G_0 \bar{\L}^{\prime}-
H^{0i} \bar{\L}^{\prime}_i,
\ee
\noindent where $\bar{\L}^{\prime}$ and $\bar{\L}^{\prime}_i$ are the
constraints $\L^{\prime}$ and $\L^{\prime}_i$ expressed in terms of the
gauge invariant combinations (45).

The equations of motion following from this Hamiltonian (46) are,
\bea
\bar {G}_0 + \fr{1}{2m}\e_{0ijk}\p^i \bar{H}^{jk}=0,\nonumber\\
\bar{G}_i + \fr{1}{2m}\e_{i\m\n\l}\p^\m \bar{H}^{\n\l}=0,\\
-\bar{H}_{0i} +\fr{1}{m}\e_{0ijk}\p^j G^k =0,\nonumber\\
-\bar{H}_{ij} +\fr{1}{m}\e_{ij\m\n}\p^\m G^\n=0.
\eea
These equations can be expressed in a covariant way, i.e.,
\bea
\bar{G}_\m + \fr{1}{2m}\e_{\m\n\l\s}{\p^\n}\bar{H}^{\l\s} = 0,\\
-\bar{H}_{\m\n} + \fr{1}{m} \e_{\m\n\l\s}\p^\l\bar{G}^\s =0.
\eea
\noindent From these equations it follows that $\p^\m \bar{G}_\m = 0$ and 
$\p^\m \bar{H}_{\m\n} = 0$. These also follow as the Hamiltonian eqations
of motion for $\bar{G}_0$ and $ \bar{H_{0i}}$, respectively.
The gauge invariant solution for 
these equations which also satisfy the divergenless
condition for $\bar{G}_\m$and $\bar{H}_{\m\n}$ are
\bea
\bar{G}_\m \equiv \tilde{H_\m} =\fr{1}{3!} \e_{\m\n\l\s}H^{\n\l\s},\nonumber\\
\bar{H}_{\m\n} \equiv  \tilde{F_{\m\n}} =\fr{1}{2}\e_{\m\n\l\s}F^{\l\s}.
\eea
\noindent where $H^{\m\n\l} =\p^\m B^{\n\l}~+~{\rm cyclic terms}$ 
and $F_{\m\n}= (\p_{\m} A_{\n} -\p_{\n} A_{\m}).$

Now, by substituting back the solutions for $\bar{G}_\m$ and $\bar{H}_{\m\n}$
in ${\cal H}_{GI}$ (46), the involutive Hamiltonian density becomes
\be
{\cal H_{GI}}=\fr{1}{4}F_{ij}F^{ij} -\fr{1}{2}F_{0i}F^{0i}
+\fr{1}{4}H_{0ij}H^{0ij} -\fr{1}{2\cdot 3!}H_{ijk}H^{ijk}+
G_0 {\tilde{\L}}-H_{0i}{\tilde{\L}^i}
\ee
\noindent where $\tilde{\L}=(\fr{1}{3!}\e_{0ijk}H^{ijk} -
\fr{1}{m}\p^i F_{0i})~~$
and$~~~\tilde{\L}^i = (-\fr{1}{2}\e_{0ijk}F^{jk} + \fr{1}{4}\p^j H_{0ij}).$

\noindent With
\bea
\fr{1}{2}\e_{0ijk}F^{jk}~=B_i,~~~~ F_{0i}~=~-E_i,\nonumber\\
\fr{1}{3!}\e_{0ijk}H^{ijk}~=~\bar{B},~~~
{\rm and}~~~ \fr{1}{2}\e_{0ijk}H^{0jk}~=~\bar{E}_i,
\eea
\noindent ${\cal H_{GI}}$ becomes,
\be
{\cal H_{GI}}=
\fr{1}{2}\left( E^2~+~B^2 \right)~+~\fr{1}{2}\left({\bar{E}}^2~
+~{\bar{B}}^2\right)~+~G_0{\tilde{\L}}-~H_{0i}{\tilde{\L}^i}
\ee

\noindent This is the Hamiltonian following from the $B{\wedge}F$ 
Lagrangian (1). Note that $\tilde{\L}~~{\rm and}~~\tilde{\L}^i $ are the
Gauss law constraints for the $B{\wedge}F$ theory. The latter, which was
an irreducible constraint in terms of gauge invariant combination$\  $
becomes a reducible constraint when expressed in terms of the solutions (51), 
obeying $\p^i{\tilde{\L}_i}=0$.
By substituting back the solutions for $\bar{G}_\m$ and 
$\bar{H}_{\m\n}$ into the equations of motion follwing from $H_{GI}$ 
(49, 50), they become
\bea
-\p^{\n} F_{\m\n} + \fr{m}{3!} \e_{\m\n\l\s} H^{\n\l\s} =0,\\
\p^{\l} H_{\m\n\l} -\fr{m}{2} \e_{\m\n\l\s}F^{\l\s} = 0 ,
\eea
\noindent which are the same equations as the one following from 
the $B{\wedge}F$ theory.
Thus under BFT embedding, the fields appearing in the original 
Hamiltonian (20) get mapped to gauge invariant combinations (45) of the
embedded Hamiltonian-$H_{GI}$ (46). The solutions to the equations 
of motion follwing from $H_{GI}$  uniquely map the embedded
Hamiltonian to that of $B{\wedge}F$ theory. Also the irreducible constraints
$\L_i~~ {\rm and}~~ \L$ in the original Hamiltonian (20) get mapped to the
reducible constraint $\tilde{\L}_i$ and $\tilde{\L}$ respectively.

\vskip0.5cm
{\large III. Phase Space Path Integral Approach}
\vskip0.5cm
The equivalence of the theory described by the Lagrangian (8) to 
$B{\wedge}F$ theory can also be established using phase space path integral
method. By suitable choice of gauge fixing conditions, the
partition function of the embedded model can  become that of original
 massive spin-1, gauge non-invariant theory or that 
of $B{\wedge}F$ theory, proving their equivalence.

The partition function for the embedded model described by the Hamiltonian
(39) is
\be
Z_{emb}~=~ \int D{\eta} \delta(\w) \delta(\th_i)\delta(\L^{\prime})
\delta(\L_i^{\prime})~\delta{(\psi_i)}~\exp~i\int {d^4x}~{\CL},
\ee
\noindent where we have omitted the trivial Fadeev-Popov determinant for
the gauges chosen below. The
 measure is,
$$ D{\eta}= D\Pi_0 D\Pi_{oi}D\Pi_{\a} D{\a} Dp_i Dq_i DG_\m DH_{\m\n},$$
and
\be
{\CL}= \Pi_0{\dot{G^0}} + \Pi_{0i}{\dot{H^{0i}}} + \Pi_{\a}{\dot{\a}}
+ p_i{\dot{q^j}} +\fr{1}{4m}\e_{0ijk}H^{ij}{\p^0 G^k} 
- \fr{1}{4m}\e_{oijk}G^i{\p^0 H^{jk}} -{\cal H_{GI}}
\ee
\noindent where ${\cal H_{GI}}$ is the invariant Hamiltonian given in (39);
$\delta{(\psi_i)}$ are the gauge fixing conditions coresponding to the 
first class constraints of the embedded model. Now we have twenty two 
independent 
phase space variables (remember $\Pi_i,~{\rm and}~\Pi_{ij}$ are not 
independent degrees of freedom) 
along with the eight first class constraints
giving the correct degrees of freedom required for 
a massive spin-one theory.

\noindent Choosing the gauge fixing conditons
\be
\delta{(\psi_i)}~=~\delta(\Pi_0)\delta(\Pi_{0i})\delta(\L)\delta(\L_i),
\ee
and integrating out the canonical conjugate variables ${\a}, \Pi_{\a}, q_i
~~ {\rm and}~~ p_i$ and momenta $\Pi_0$ and $\Pi_{0i}$ from the partition
function reduces $Z_{emb}$ to 
\be
Z~=~\int DG_\m DH_{\m\n} \delta(\L) \delta (\L_i) exp~i~\int {d^4x}~{\CL},
\ee
\noindent where ${\CL}$ is the original first order Lagrangian (8), with the Gauss law constraints
impossed through $\delta{(\L)}$ and $\delta{(\L_i)}.$ Note that the
original second class constraints are the gauge fixing conditions (59).

Next we choose the gauge fixing conditions as
\be
\delta{(\psi_i)}~=~\delta(\p^i H_{0i}) \delta(\p^iq_i) \delta{(\chi_i)}
\delta{(\chi_{ij})},
\ee
\noindent where
\bea
\chi_i = (G_i~-~\fr{1}{m}\e_{0ijk}\p^j H^{0k})\\
\chi_{ij} = (H_{ij}~-~\fr{1}{m}\e_{0ijk}\p^k G^0),
\eea
\noindent to show the equivalence of the embedded model to the $B{\wedge}F$
theory.
Owing to the constraints $\w~{\rm and}~\th_i$, the $D\Pi_0,~D\Pi_{0i},$
 integrations are trivial. The $D{\Pi_\a},~{\rm and}~D p_i$
integrations along with the constraints $\delta(\L^{\prime}),~{\rm and}~
\delta(\L^{\prime}_i)$ lead to the terms $-\fr{1}{4 m^2} G_{ij}G^{ij}$ and 
$~\fr{1}{2\cdot 3!} H_{ijk}H^{ijk}$ in the exponent. Using the fact 
that $\p^i G_i=0$ and $\p^i H_{ij}=0$ on the constraint surface of 
$\chi_i$ and $\chi_{ij}$ and the gauge fixing conditions $\delta(\p^iq_i)$
and $\delta(\p^i H_{0i})$, we carry out
the integrations over $Dq_i$ and $D\a$, which are just Gaussian. Thus the
Lagrangian in the partition function becomes,
\bea
{\CL}=\fr{1}{2m}\e_{0ijk}\p^0 G^i H^{jk} -\fr{1}{4m^2} G_{ij}G^{ij} 
+\fr{1}{2\cdot 3! m^2} H_{ijk}H^{ijk} -\fr{1}{2m^2}G_{0i}G^{0i} +
\fr{1}{4m^2} H_{0ij}H^{0ij} \nonumber\\
\!\!\!\!\!\!\!\!\!\!\!\!\!\!\!\!\!\!\!\!\!\!\!\!\!\!\!\!\!\!\!\!\!\!\!\!\!
\!\!\!\!\!\!\!\!\!\!\!\!\!\!\!\!\!\!\!\!\!\!\!\!\!\!\!\!\!\!\!\!\!\!\!\!\!
\!\!\!\!\!\!\!\!\!\!\!\!\!\!\!\!\!\!\!\!\!\!\!\!\!\!\!\!\!\!\!\!\!\!\!\!\!
\!\!\!\!\!\!\!\!\!\!\!\!\!\!\!\!\!\!\!\!\!\!\!\!\!\!\!\!\!\!\!\!\!\!\!\!\!
\!\!\!\!\!\!\!\!\!\!\!\!\!\!\!\!\!\!\!\!\!\!\!\!\!\!\!\!\!\!\!\!\!\!\!\!\!
\!\!\!\!\!\!\!\!\!\!\!\!\!\!\!\!\!\!\!\!\!\!\!\!\!\!\!\!\!\!\!\!\!\!\!\!\!
\!\!\!\!\!\!\!\!\!\!\!\!\!\!\!\!\!\!\!\!\!\!\!\!\!\!\!\!\!\!\!\!\!\!\!\!\
 -\fr{1}{4}H_{ij}H^{ij}
+\fr{1}{2} G_i G^i
-\fr{1}{2m^2}H_{0i}{\nabla}^2H^{0i} +\fr{1}{2m^2} G_0{\nabla}^2 G^0~.
\eea
\noindent After
using the constraints (62, 63) and the conditions 
$\nabla^2 H_{0i}=-\fr{m}{2}\e_{0ijk}G^{jk}$ and
$\nabla^2 G_0=~\fr{m}{2} \e_{0ijk}\p^i H^{jk}$ implied by (62, 63),
the partition function becomes,
\be
Z~=~ \int DA_\m DB_{\m\n}\delta{(\chi_i)}
\delta{(\chi_{ij})}
~exp~i~\int {d^4x}~{\CL}
\ee
\noindent where ${\CL}$, with the  identifications
\bea
\fr{1}{m}G_\m = A_\m,\\
\fr{1}{m}H_{\m\n}= B_{\m\n}
\eea
 is the Lagrangian of $B{\wedge}F$ theory(18).
The constraints $\chi_i ~~ {\rm and} 
~~ \chi_{ij}$, in terms of $A_\m ~{\rm and}~ B_{\m\n}$ become,
\bea
\chi_i= (m A_i - \e_{0ijk}\p^jB^{0k}),\\
\!\!\!{\rm and} ~~~~
\chi_{ij} = (m B_{ij} -\e_{0ijk}\p^k A^0)
\eea
$\chi_{i}$ and $\chi_{ij}$ which are the gauge
fixing conditions for the linearly independent generators $\th_i$ and 
$\L^\prime_i,$
now play the role of Gauss law constraints of the $B{\wedge}F$ theory.
Now 
\bea
\nabla^2B_{0i}=-\fr{m}{2}\e_{0ijk}F^{jk},\\
{\rm and}~~~~~
\nabla^2 A_0 = \fr{m}{2}\e_{0ijk}\p^i B^{jk},
\eea
\noindent ($F_{jk} = \p_j A_k-\p_kA_j$), are the Gauss law constraints in 
the gauge $\chi_i=0$ and $\chi_{ij}=0$.
Since $\th_i$ is not a reducible constraint, coresponding gauge fixing 
condition
is also not reducible; but it implies the reducible Gauss law constraint (70)
present in the $B{\wedge}F$ theory. It is interesting to note the 
complimentary behavior of Gauss law constraints which comes as the gauge fixing
conditions in the partition function for the embedded model.
\vskip0.5cm
{\large Conclusion}
\vskip0.5cm
In this paper we have started with a new first order formulation of
massive spin-one theory which is gauge non-invariant and converted it to a
theory with only first class constraints following the line of 
Hamiltonian embedding of BFT. We showed that the embedded Hamiltonian 
is equivalent to the Hamiltonian of $B{\wedge}F$ theory. This was shown
both from the solutions of equations of motion following from the embedded
Hamiltonian and from the phase space path integral in a suitable gauge. We
also point out how an irreducible constraint of the first order theory
gets mapped to the reducible constraint of the $B{\wedge}F$ theory.
It should be pointed out that the first-order lagrangian (8) and its 
equivalence to topologically massive gauge theory are both new results.

A similar first order Lagrangian can be formulated for massive spin-zero
particle, but now involving a 3-form and a scalar fields as
\be
{\CL} = -\fr{1}{2\cdot 3!} C_{\m\n\l}C^{\m\n\l} -\fr{1}{2}{\phi}{\phi}
+\fr{1}{4!~m}\e_{\m\n\l\s}C^{\m\n\l\s}{\phi},
\ee
\noindent where $C_{\m\n\l\s} = \p_\m C_{\n\l\s}~+~{\rm cyclic~terms}.$
Interestingly here the field content is the same as that of DKP formulation
of spin-zero theory.

The equivalence demonstrated here is of the same nature as that between 
self-dual model \cite{dj} and Maxwell-Chern-Simmon theory in $2+1$ dimensions, 
shown in \cite{rb1}.
The behavior of the fields of the embedded Hamiltonian here is the same
as that of $2+1$ dimension self-dual model; viz, the gauge variant fields
of the embedded model can be mapped to the fundamental fields of the $B{\wedge}F$ theory or
that of the original model. Despite this similarity, model in (8) is
different from the self-dual model. The latter describes only half the degrees
of freedom compared to that of massive spin-one theory in $2+1$ dimensions
and consequently is equivalent to the parity violating 
Maxwell-Chern-Simmon theory. Also the self-duality condition is possible
only in $4k-1$ dimensions. But the former describes all the three 
states of polarization needed for massive spin-one particle and also this
construction is possible in all dimensions and has a even-parity mass term.
Owing to the even-parity mass term described by this model, the $2+1$
dimensional non-abelian generalization of (8) may be related to the recently 
constructed Jackiw-Pi model \cite{jp}. The self-dual model and 
Maxwell-Chern-Simmon correspondence has proved to be useful in
Bosonization in $2+1$ dimensions \cite{rb5}. It should be interesting to see
similarly if the equivalence proved here has a role in studying Bosonization
in $3+1$ dimensions. It is also interesting to study
the Hamiltonian embedding of the non-abelian version of (8). 
Work along these lines are in progress.
  
  We have exploited the gauge symmetry arising due to the Hamiltonian
embedding procedure to prove the equivalence between two different formulations
of massive spin-one theory. There is a different procedure which also has
the potential to establish equivalence among different formulations 
\cite{dam}. In this method, the new degrees of freedom are added by hand,
which generates abelian gauge algebra and this is used to gauge fix
suitably, to arrive at a different formulation of the original theory
like for example Bosonisation \cite{smooth}.
It should be interesting to investigate if the quantum equivalence 
proved here survives the
case of coupling with external fields like gravitation and 
electromagnetism also.
\vskip0.5cm
\noindent {\bf Acknowledgements}:
We are grateful to Prof. V. Srinivasan for pointing out an error in the earlier
manuscript. We acknowledge Dr. R. Banerjee for useful corespondence. We
thank the referee for bringing the reference \cite{dam} to our notice.
 EH thanks 
U.G.C., India for support through S.R.F scheme.

\end{document}